\documentclass[a4paper,11pt]{article}
\usepackage{jinstpub}   % JINST style: page layout, title page, fonts, natbib, hyperref

\usepackage{graphicx}
\usepackage{booktabs}
\usepackage{multirow}
\usepackage{amsmath}
\usepackage{tabularx}
\usepackage{xcolor}
\usepackage{textcomp}
\usepackage{microtype}
\title{A Reproducible Benchmark and Evidence-Retrieval Software Framework for Silicon Detector R\&D Literature}

\author[a]{Tianqi Gao,}
\author[b,1]{Ruobing Jiang,\note{Corresponding author.}}
\author[b]{Dawei Fu,}
\author[b]{Qiang Li}
\author[a]{and Matthew Kenzie}

\affiliation[a]{Cavendish Laboratory, University of Cambridge,\\
  Cambridge CB3 0US, United Kingdom}
\affiliation[b]{State Key Laboratory of Nuclear Physics and Technology, Peking University,\\
  Beijing 100871, China}

\emailAdd{ruobing@stu.pku.edu.cn}

\abstract{Silicon pixel detector R\&D depends on a large and rapidly growing technical literature, including beam-test and irradiation studies, performance measurements, simulation, and design reports. Locating the supporting evidence passage for a measurement, operating condition, or design decision is therefore a computing and data-science challenge for detector-development workflows. General-purpose language models are insufficient unless grounded in traceable primary sources, particularly in a domain with specialised terminology, configuration-dependent measurements, and rapidly evolving experimental results. We address this with a reproducible, general-purpose framework for evidence-grounded retrieval over technical literature, using silicon pixel detector R\&D as a demanding validation domain. The framework combines sparse lexical retrieval, dense semantic retrieval, and hybrid reciprocal-rank fusion, with an optional graph-guided exploration layer and grounded, abstention-aware response generation. The accompanying benchmark provides manually curated chunk-level evidence annotations, source-level diagnostics, semantic relevance checks, and negative-query abstention tests over two detector query sets. We evaluate six retrieval configurations across 378 source documents and 8,442 indexed chunks. Hybrid sparse-dense retrieval gives the strongest strict evidence recovery, achieving Hit@5 of 0.917 on the core benchmark and 0.951 on the curated extension benchmark, while graph-based methods are more effective for literature exploration and source discovery. Graph expansion is therefore best employed as a discovery layer over the hybrid retrieval backbone. The framework provides reusable software for traceable, evidence-grounded knowledge access in silicon detector R\&D and high-energy physics instrumentation.}


\arxivnumber{2606.24725}

\begin{document}
\maketitle
\flushbottom

\section{Introduction}
\label{sec:intro}

Silicon pixel detectors form the tracking and vertexing core of modern particle-physics experiments, and their performance directly shapes the physics an experiment can deliver. Spatial resolution, radiation tolerance, and timing govern how precisely charged particles are reconstructed, how efficiently heavy-flavour decays are tagged, and how robustly events are resolved under high pile-up, and these capabilities in turn set the reach of measurements ranging from precision Higgs and flavour physics to searches for new phenomena. The detector-design and radiation-qualification choices that determine this performance are therefore not purely technical, and they define the physics reach of an experiment. This is especially true for the high-luminosity LHC upgrades and for proposed future colliders, where new silicon detector systems must be designed and qualified by drawing together quantitative evidence accumulated across many sensor technologies, irradiation campaigns, beam tests, and simulation studies. Reliable and traceable access to that distributed evidence base is thus a capability that directly supports the underlying physics programme, and it is the problem this work addresses. We approach this as a general problem in evidence-grounded literature retrieval and present a reproducible framework that applies retrieval-augmented generation (RAG)~\cite{lewis2020rag} to a specialised technical corpus, using silicon pixel detector R\&D as a demanding validation domain rather than the sole target of the method.

This research spans sensor design, radiation qualification,
beam-test characterisation, simulation, and electronics integration, with
individual studies drawing on information across detector technologies,
experimental conditions, and software frameworks. This breadth is scientifically valuable, but it also creates a practical problem for detector R\&D: the evidence needed to justify a detector-design choice or performance claim is often distributed across many publications and subfields. Document-level search is therefore frequently insufficient. For example, a detector study may require a specific passage on how substrate bias affects charge-collection efficiency after irradiation, the timing performance achieved for a particular LGAD geometry~\cite{pellegrini2014lgad}, or the threshold and
reconstruction conditions used in a beam test. Similarly, questions involving High-Voltage Monolithic Active Pixel Sensor (HV-MAPS), Depleted Monolithic Active Pixel Sensor (DMAPS), or hybrid pixel assemblies~\cite{snoeys2014maps,pernegger2017hvcmos} may require distinguishing sensor architectures, irradiation fluences, operating voltages, and readout configurations that are superficially similar but physically distinct.
Large language models (LLMs) are increasingly adopted for scientific question answering; however, they remain limited in generating faithful and evidence-grounded responses for specialised detector knowledge, long-tail technical details, and rapidly evolving detector developments~\cite{ji2023survey,gao2023ragsurvey,mallen2023trust}. Many detector-R\&D questions depend on information that appears only in a small number of publications and under specific operating conditions, making such knowledge difficult to recover reliably from the parametric memory of a language model alone. RAG mitigates this limitation by grounding answer generation in evidence retrieved from external corpora. 

In practice, the retrieval step is itself non-trivial for detector literature. The challenge is not simply retrieving related publications, but identifying the exact supporting evidence under the relevant detector configuration, irradiation condition, bias voltage, or readout setting. Dense semantic retrieval enables matching of paraphrased queries~\cite{reimers2019sbert,karpukhin2020dpr,thakur2021beir}, but may miss exact acronyms, detector names, or process-specific terminology. Sparse lexical retrieval such as BM25 remains a strong baseline for keyword-driven queries~\cite{robertson2009bm25,manning2008ir}. Hybrid approaches, for example using reciprocal-rank fusion (RRF)~\cite{cormack2009rrf,lin2021sparse}, combine these signals and often provide robust performance across heterogeneous query types. Recent graph-guided retrieval approaches further introduce structured relationships between entities, concepts and documents to support broader literature exploration~\cite{edge2024graphrag,guo2024lightrag,yasunaga2021qagnn,xiong2021multihop}. While these methods can improve semantic exploration and literature navigation, they are not automatically beneficial for detector-domain retrieval: graph expansion and semantic broadening can introduce passages that are topically related but less directly relevant to the required detector evidence. This tension between semantic coverage and strict evidence precision is particularly important for detector R\&D, where an unsupported or weakly grounded statement, if taken as established, can propagate into detector-design choices or physics-analysis decisions.

\begin{figure*}[t]
\centering
\includegraphics[width=0.95\textwidth]{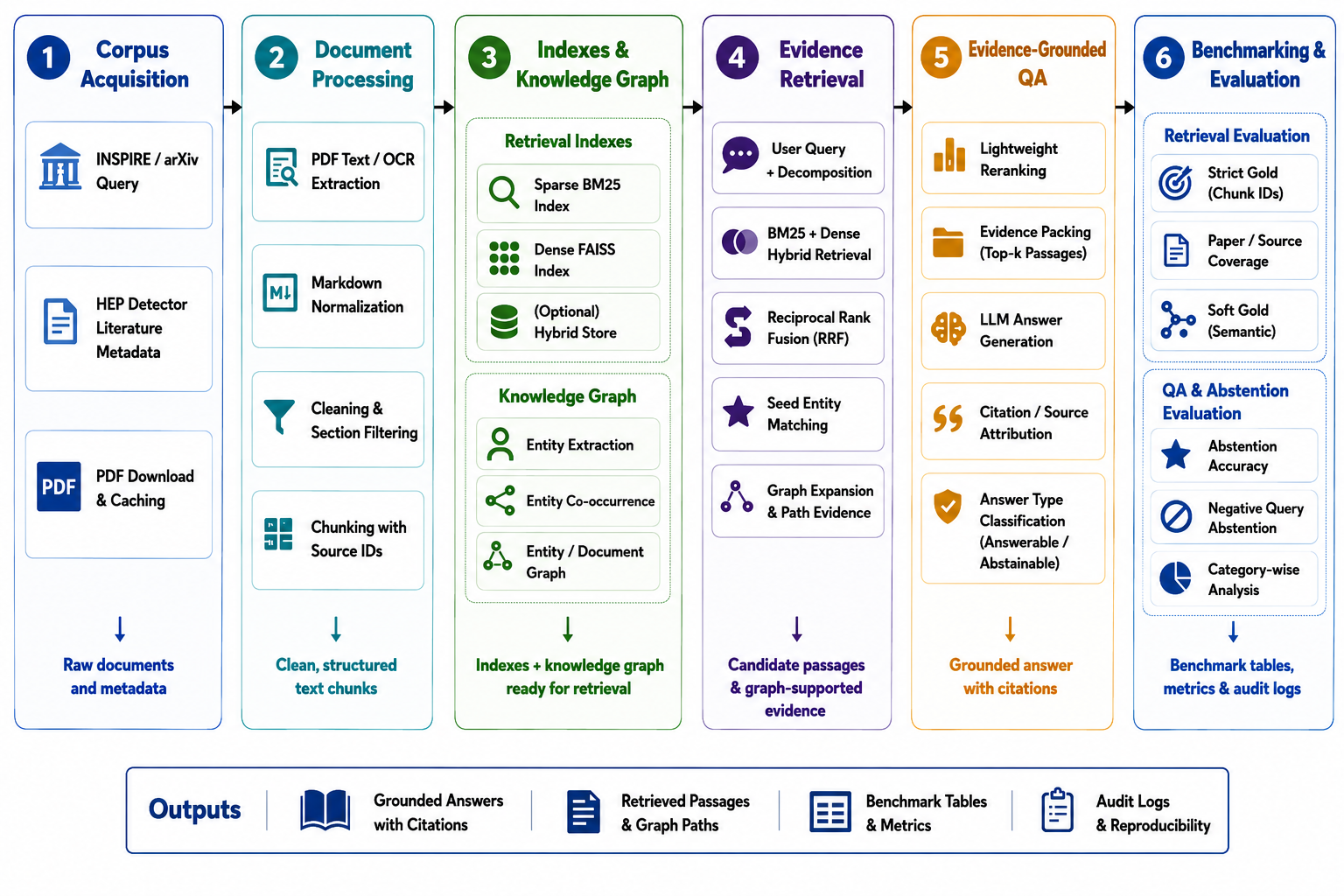}
\caption{Overview of the complete system pipeline. The six stages cover corpus
acquisition, document processing, index and knowledge-graph construction, hybrid and
graph-guided retrieval, grounded answer generation, and rigorous evaluation and
benchmarking.}
\label{fig:pipeline}
\end{figure*}

In this work, we instantiate and evaluate this framework for silicon pixel detector literature, framed as a software and computing tool for detector R\&D rather than as a general NLP benchmark. Rather than proposing a new retrieval algorithm, the aim of this study is to understand how established retrieval strategies behave in a detector-instrumentation setting. By using standard and well-understood retrieval components, we isolate the domain effect and directly examine whether graph-enhanced retrieval improves strict evidence ranking for detector literature, or whether its primary value lies instead in literature exploration and detector-concept navigation. We therefore combine BM25 sparse retrieval, dense FAISS retrieval, reciprocal-rank-fusion hybrid retrieval, detector-entity graph expansion, and graph-path retrieval within a unified evaluation pipeline, summarised in Figure~\ref{fig:pipeline}. Retrieval performance is evaluated using manually curated chunk-level gold evidence, source-level diagnostics, semantic soft-gold checks, and negative-query abstention tests across two complementary detector-domain benchmarks. The framework's role is to surface and rank traceable supporting evidence, and to abstain when the corpus contains none. Interpreting that evidence and making detector-design decisions remains with the human domain expert. We therefore evaluate the framework on whether it retrieves the correct supporting passage and abstains appropriately, not on end-to-end decision quality. The results show that hybrid sparse--dense retrieval provides the strongest strict evidence-ranking performance, while graph-based approaches remain valuable for exploratory literature navigation and detector-concept interpretation. 

More broadly, this work sits within the growing effort to develop
computing, software, and data-science tools for high-energy physics.
Reliable, source-traceable access to a large and heterogeneous technical
literature is itself part of the software and data infrastructure that
supports detector R\&D, alongside the simulation, reconstruction, and
machine-learning tools more commonly associated with the field. Recent
work has begun to apply retrieval-augmented and language-model methods to
specialised high-energy-physics corpora~\cite{Jiang:2026pkh}, and we
contribute to this direction a reproducible, evaluation-driven framework
together with an openly released benchmark, so that retrieval quality for
scientific evidence access can be measured rather than assumed. We treat
the framework and benchmark as research-software artefacts: the
corpus-processing, indexing, retrieval, and evaluation components are
released to support reproduction, auditing, and adaptation to other
detector-literature corpora.

The main contributions of this work are:
\begin{itemize}
\item the first detector-domain benchmark for grounded evidence retrieval in silicon pixel detector literature, where success is defined as retrieving the exact supporting passage for a detector claim (or correctly abstaining when none exists), supported by strict chunk-level annotations, source-level diagnostics, semantic soft-gold evaluation, and negative-query abstention tests;

\item a reproducible retrieval framework combining BM25 sparse retrieval, dense FAISS retrieval, hybrid reciprocal-rank-fusion retrieval, detector-entity graph expansion, and graph-path retrieval under a unified evaluation pipeline;

\item an empirical study of retrieval behaviour in detector instrumentation literature, showing that hybrid sparse--dense retrieval provides the strongest strict evidence-ranking performance, while graph-based retrieval is more effective for literature exploration and detector-concept navigation than for exact evidence retrieval, providing a practical domain-specific caution against adopting GraphRAG-style expansion as a default evidence ranker;

\item detector-facing benchmark queries and retrieval analyses covering irradiation studies, timing performance, charge collection, detector simulation, readout configuration, and detector-technology comparison, demonstrating how reliable evidence retrieval can support, rather than replace, expert judgement in silicon detector R\&D.
\end{itemize}
\section{Background and Motivation}
\label{sec:background}

\subsection{Scientific retrieval and evidence-grounded literature access}

Detector-physics literature differs from generic open-domain corpora in both language and evidential structure. Terminology is dense with acronyms, device names, detector design-specific concepts, and configuration-dependent measurements that often carry highly specialised meanings. Different claims within the same paper may depend on distinct irradiation conditions, bias voltages, ASIC configurations, beam-test assumptions, or reconstruction settings, and the supporting evidence therefore takes the form of specific evidence fragments rather than entire documents. These characteristics make detector literature a long-tail scientific knowledge domain in which critical design and performance information is distributed across specialised publications rather than concentrated in a small number of widely cited sources. Consequently, detector-domain retrieval differs fundamentally from generic document search and motivates the use of evidence-grounded retrieval pipelines.

RAG grounds downstream analysis and answer generation in retrieved evidence passages rather than relying entirely on parametric language-model memory~\cite{lewis2020rag,izacard2021fid}. Recent work has demonstrated the value of evidence-grounded retrieval in specialised high-energy-physics domains, including muon-collider phenomenology and future-collider studies~\cite{Jiang:2026pkh}. Hybrid sparse--dense retrieval combines lexical precision with semantic matching and is widely used in scientific and technical retrieval settings~\cite{karpukhin2020dpr,khattab2020colbert}. Graph-based retrieval and entity-linking approaches, including GraphRAG-style methods, can additionally expose relationships between detector technologies, irradiation studies, readout architectures, and performance measurements, providing useful contextual expansion and interpretability~\cite{edge2024graphrag}. However, broader semantic expansion is not automatically beneficial for scientific retrieval. Retrieving loosely related passages can weaken evidence precision if the highest-ranked citations become detached from the detector configuration or measurement context relevant to the original query. This trade-off between semantic coverage and evidence precision is particularly important for detector R\&D, where unsupported or weakly grounded statements may influence detector-design choices, performance projections, or technology-selection decisions. This consideration becomes increasingly relevant for future collider programmes, where detector concepts must be evaluated using evidence accumulated across multiple sensor technologies, irradiation campaigns, beam tests, and simulation studies.

Several benchmark datasets exist for scientific retrieval and evidence-grounded information access, including SciFact~\cite{wadden2020scifact}, SciDocs~\cite{cohan2020specter}, and biomedical retrieval benchmarks such as BioASQ~\cite{tsatsaronis2015bioasq}. However, these benchmarks target generic scientific or biomedical literature and do not capture the domain-specific characteristics of silicon pixel detector publications: configuration-dependent measurements, irradiation conditions, exact detector terminology, and passage-local evidence under stated operating conditions. A detector-domain benchmark therefore requires dedicated corpus construction, query design, and chunk-level gold annotation that cannot be substituted by existing general-purpose scientific retrieval resources.

\subsection{Retrieval challenges in silicon detector literature}

To make the retrieval challenge concrete, consider a representative detector-R\&D query: \textit{``What is the time resolution of LGAD sensors after being irradiated by a fluence of $10^{15}$~n$_{\rm eq}$/cm$^2$?''} Answering this query correctly requires identifying publications reporting LGAD test results after neutron irradiation, matching the fluence level to the relevant experimental conditions, distinguishing irradiated from unirradiated timing performance, and verifying whether parameters such as bias voltage, temperature, sensor thickness, and gain-layer design are stated. Similar challenges arise for questions involving HV-MAPS, DMAPS, hybrid pixel sensors, detector simulations, or irradiation studies, where seemingly similar measurements may correspond to substantially different detector configurations.

Detector-physics reasoning of this kind also requires understanding that many entities are physically related. A query about depletion voltage may imply relevance to electric field formation, charge collection, sensor capacitance, and bias optimisation. Likewise, a query concerning timing performance may require connecting sensor geometry, gain mechanisms, front-end electronics, and irradiation behaviour. This relational structure is one reason why graph-guided retrieval and entity linking can provide useful contextual expansion. However, graph expansion can also introduce adjacent but non-gold passages that pass semantic soft-gold checks while failing strict evidence matching. In practice, graph-expanded context is valuable for navigating detector-concept relationships and recovering surrounding technical context, but it should not displace precision retrieval when an exact supporting passage exists in the corpus. This distinction becomes increasingly important as detector studies grow in scale and complexity, requiring evidence to be integrated across multiple sensor technologies and experimental campaigns.

\subsection{Application scenarios for detector R\&D}
\label{sec:applications}

To characterise the structure of information needs in silicon pixel detector R\&D, we define representative application scenarios for the proposed framework in Table~\ref{tab:applications}. These scenarios span detector-technology comparison, irradiation and radiation-hardness studies, timing-performance characterisation, charge-collection and signal-formation analysis, readout and front-end configuration, detector-simulation and test-beam workflows, and answerability checking. Such information-access requirements are expected to become increasingly important for future collider detector programmes, where detector-design decisions must be informed by evidence accumulated across multiple sensor technologies, irradiation campaigns, beam tests, and simulation studies. Each scenario maps a recurring detector-R\&D question to its underlying retrieval objective, motivating a system capable of chunk-level evidence retrieval, source traceability, and evidence-grounded literature access under stated operating conditions.

\begin{table*}[!htbp]
\centering
\caption{Representative retrieval application scenarios and query intents for silicon pixel detector R\&D.}
\label{tab:applications}
\begin{tabularx}{\linewidth}{p{0.22\linewidth} X p{0.27\linewidth}}
\toprule
\textbf{Query category} & \textbf{User intent} & \textbf{Retrieval objectives} \\
\midrule
Technology comparison &
How do HV-MAPS, DMAPS, hybrid pixels, and LGAD-based detectors compare for a given application? &
Sensor-architecture and performance trade-off studies. \\

Irradiation \& radiation hardness &
How does charge collection or leakage current evolve with fluence and bias? &
Radiation-qualification and annealing studies under stated fluence. \\

Timing performance &
What time resolution is achieved for a given LGAD or AC-LGAD geometry? &
Beam-test and laboratory timing characterisation. \\

Charge collection \& signal formation &
How do depletion, electric field, and weighting field shape the collected signal? &
Sensor-physics and TCAD/charge-transport studies. \\

Readout \& front-end &
What threshold, ToA/ToT, and calibration conditions are used? &
ASIC, Timepix, and front-end configuration reports. \\

Simulation \& test beam &
Which simulation framework and reconstruction setup were used? &
Allpix Squared, Geant4, and test-beam analysis literature. \\

Answerability check &
Is a requested detector claim supported by the available corpus? &
Abstain rather than produce unsupported answers. \\
\bottomrule
\end{tabularx}
\end{table*}

These characteristics create a practical challenge for literature navigation. Detector-R\&D questions often require connecting sensor architecture, operating conditions, irradiation behaviour, and readout configuration across multiple sources. A query on the timing resolution of an irradiated LGAD, for instance, requires different evidence from one on charge-collection efficiency in a DMAPS sensor, even though both arise within the same corpus. This motivates a retrieval framework capable of chunk-level evidence retrieval, source traceability, and evidence-grounded literature access for silicon pixel detector studies.

\section{Methodology}
\label{sec:method}

\subsection{Design principles and corpus construction}
\label{sec:corpus}

Figure~\ref{fig:pipeline} summarises the complete framework. The system is designed to support detector-oriented workflows involving irradiation studies, timing measurements, charge-collection analysis, detector simulation, sensor-technology comparison, and detector-performance interpretation. Unlike generic scientific retrieval tasks, detector-domain retrieval presents three distinctive challenges. First, supporting evidence is frequently passage-level rather than document-level, with critical detector-performance information often appearing only within a small portion of a publication. Second, detector claims depend strongly on operating conditions such as irradiation fluence, bias voltage, temperature, particle type, and readout configuration. Third, many detector-R\&D questions involve long-tail technical knowledge distributed across specialised publications and detector-specific studies. These characteristics motivate a retrieval-first architecture that prioritises evidence precision, source traceability, and detector-configuration awareness.

The retrieval framework therefore combines a high-precision hybrid retrieval backbone with a complementary graph-guided exploration layer. Hybrid retrieval is intended to maximise strict evidence recovery for detector-R\&D questions, while graph-guided expansion provides controlled contextual exploration of related detector concepts and technologies without displacing the strongest first-stage retrieval results. This separation allows the framework to support both evidence-focused retrieval and broader literature navigation.

The corpus consists of silicon detector instrumentation literature and associated software resources, including journal articles, conference proceedings, technical design reports, detector-simulation documentation, and detector-development studies. It is assembled by automated keyword search rather than manual curation, as detailed in Section~\ref{sec:Experimental setup}. The resulting documents frequently report quantitative measurements under stated operating conditions, including irradiation fluence, bias voltage, temperature, timing performance, charge collection, detector geometry, and readout configuration.

Source documents are converted into cleaned text using standard PDF extraction followed by lightweight preprocessing~\cite{lopez2009grobid}. Headers, footers, watermarks, and duplicated captions are removed where possible, while section headings, mathematical expressions, physical units, detector terminology, and numerical values are preserved because they frequently carry the experimental context required to interpret detector-performance measurements. Documents are segmented into overlapping passage-level chunks of approximately 300--500 tokens, with section boundaries preserved where possible so that measurement conditions remain associated with the corresponding quantitative results.

Each chunk stores hierarchical identifiers and source metadata linking it to the original document and publication context. The final indexed corpus comprises 378 source published documents segmented into 8,442 passage-level chunks. These identifiers support both strict chunk-level evaluation and source-level retrieval analysis, allowing exact evidence retrieval to be distinguished from retrieval of the correct publication without the supporting passage.

\subsection{Precision-oriented evidence retrieval}
\label{sec:retrieval}

Because detector-performance claims often depend on exact operating conditions, first-stage retrieval must maximise evidence precision while remaining robust to paraphrased detector terminology. A retrieval system that retrieves the correct publication but fails to recover the supporting passage is insufficient for evidence-grounded detector studies. We therefore employ a hybrid retrieval strategy that combines sparse lexical matching with dense semantic retrieval. Sparse retrieval preserves exact detector terminology and technical acronyms, while dense retrieval improves robustness to paraphrased detector-physics descriptions. The two signals are subsequently fused to provide a high-precision evidence-retrieval backbone for detector-domain literature search.

\paragraph{Sparse retriever.}
The sparse component uses BM25~\cite{robertson2009bm25}, which scores a tokenized
query $q$ and document chunk $d$ according to term frequency, inverse document
frequency, and length normalization:
\begin{equation}
S_{\rm BM25}(q,d)=
\sum_{t\in q}
\mathrm{IDF}(t)\,
\frac{f(t,d)(k_1+1)}
{f(t,d)+k_1\!\left(1-b+b\,|d|/\overline{|d|}\right)},
\end{equation}
where $f(t,d)$ is the frequency of term $t$ in chunk $d$, $|d|$ is the chunk length,
$\overline{|d|}$ the mean chunk length, and $k_1$ and $b$ are free parameters. The
inverse document frequency $\mathrm{IDF}(t)=\log\frac{N}{df(t)}$ assigns higher
weight to rare terms, where $N$ is the total number of chunks and $df(t)$ the number
containing $t$. We use $b=0.75$ as a standard length-normalization compromise. BM25
prioritises exact term overlap, making it particularly effective in detector
literature where key concepts are expressed through stable acronyms and technical
keywords such as LGAD, DMAPS, HV-MAPS, ToA/ToT, and $\mathrm{n}_{\rm eq}/\mathrm{cm}^2$.
Its main limitation is that it cannot capture semantic equivalence when relevant
concepts are expressed using different terminology, motivating the complementary use
of dense retrieval.

\paragraph{Dense retriever.}
The dense component embeds queries and chunks into a shared vector space using
\texttt{sentence-\allowbreak transformers/\allowbreak all-\allowbreak MiniLM-\allowbreak L6-\allowbreak v2}, indexed with FAISS under cosine
similarity~\cite{johnson2019faiss}. Let $f(\cdot)$ denote the encoder mapping a text
input to an embedding vector, $\mathbf{e}_q=f(q)$ and $\mathbf{e}_d=f(d)$. Dense
retrieval ranks chunks by cosine similarity:
\begin{equation}
S_{\rm dense}(q,d)=
\frac{\mathbf{e}_q\cdot\mathbf{e}_d}
{\|\mathbf{e}_q\|\,\|\mathbf{e}_d\|},
\qquad
d^{*}=\arg\max_{d\in\mathcal{D}} S_{\rm dense}(q,d).
\end{equation}
Dense retrieval captures conceptual similarity even when surface forms differ, for
example matching ``charge-collection efficiency'' to ``collected-charge fraction''.
However, embedding-based similarity can over-generalise on fine-grained technical
queries and may underperform on rare acronyms or newly introduced detector
terminology, which motivates combining it with BM25 in a hybrid scheme.

\paragraph{Hybrid retriever.}
The two ranked lists are merged using weighted reciprocal-rank fusion (RRF):
\begin{equation}
S_{\rm RRF}(c)=\frac{w_d}{K+r_d(c)}+\frac{w_b}{K+r_b(c)},
\label{eq:rrf}
\end{equation}
where $r_d(c)$ and $r_b(c)$ denote the dense and BM25 ranks of chunk $c$, and $w_d$
and $w_b$ are the corresponding fusion weights. We set $K=60$ following the original
RRF formulation~\cite{cormack2009rrf}, which acts as a smoothing term that stabilises
fusion across retrievers. The default values $w_d=0.9$ and $w_b=0.1$ are selected
based on a grid search over weight combinations, as described in
Section~\ref{sec:experiments}. Hybrid retrieval preserves the complementary strengths
of lexical and semantic retrieval: BM25 is effective for acronym-specific detector
queries, while dense retrieval improves robustness to paraphrased detector-physics
descriptions and semantically related terminology.

\subsection{Graph-guided literature exploration}
\label{sec:graph}

Graph-guided expansion is evaluated as a complementary literature-exploration strategy rather than a replacement for high-precision evidence retrieval. Detector entities and concepts are linked through a lightweight graph representation constructed from co-occurrence and metadata relationships across the indexed corpus. The graph covers detector technologies, sensor-physics concepts, readout electronics, performance metrics, irradiation studies, and simulation frameworks. Starting from the initial hybrid-retrieval results, graph traversal is used to identify neighbouring detector concepts and related passages that may provide additional contextual information.

In addition to static graph expansion, we evaluate an agentic graph configuration
that introduces a lightweight query-decomposition stage before retrieval.
Complex detector-domain questions are first decomposed into a small set of
sub-queries targeting detector technologies, performance metrics, operating
conditions, or sensor-physics concepts. Retrieval is then performed over the
graph-expanded search space and the resulting evidence is merged before final
ranking. The purpose of this configuration is not to optimise strict evidence
retrieval, but to investigate whether structured decomposition can improve
coverage for multi-concept detector questions involving several interacting
entities or operating conditions.

The primary purpose of graph-guided exploration is to support literature navigation and detector-concept discovery. Detector-R\&D questions frequently involve relationships between multiple technologies, operating conditions, and performance metrics that are not always captured by direct lexical matching. Graph-based exploration can therefore help expose connections between detector technologies, irradiation behaviour, charge-collection mechanisms, timing-performance studies, and simulation frameworks, providing broader contextual understanding of the retrieved evidence.

\begin{figure*}[t]
\centering
\includegraphics[width=\textwidth]{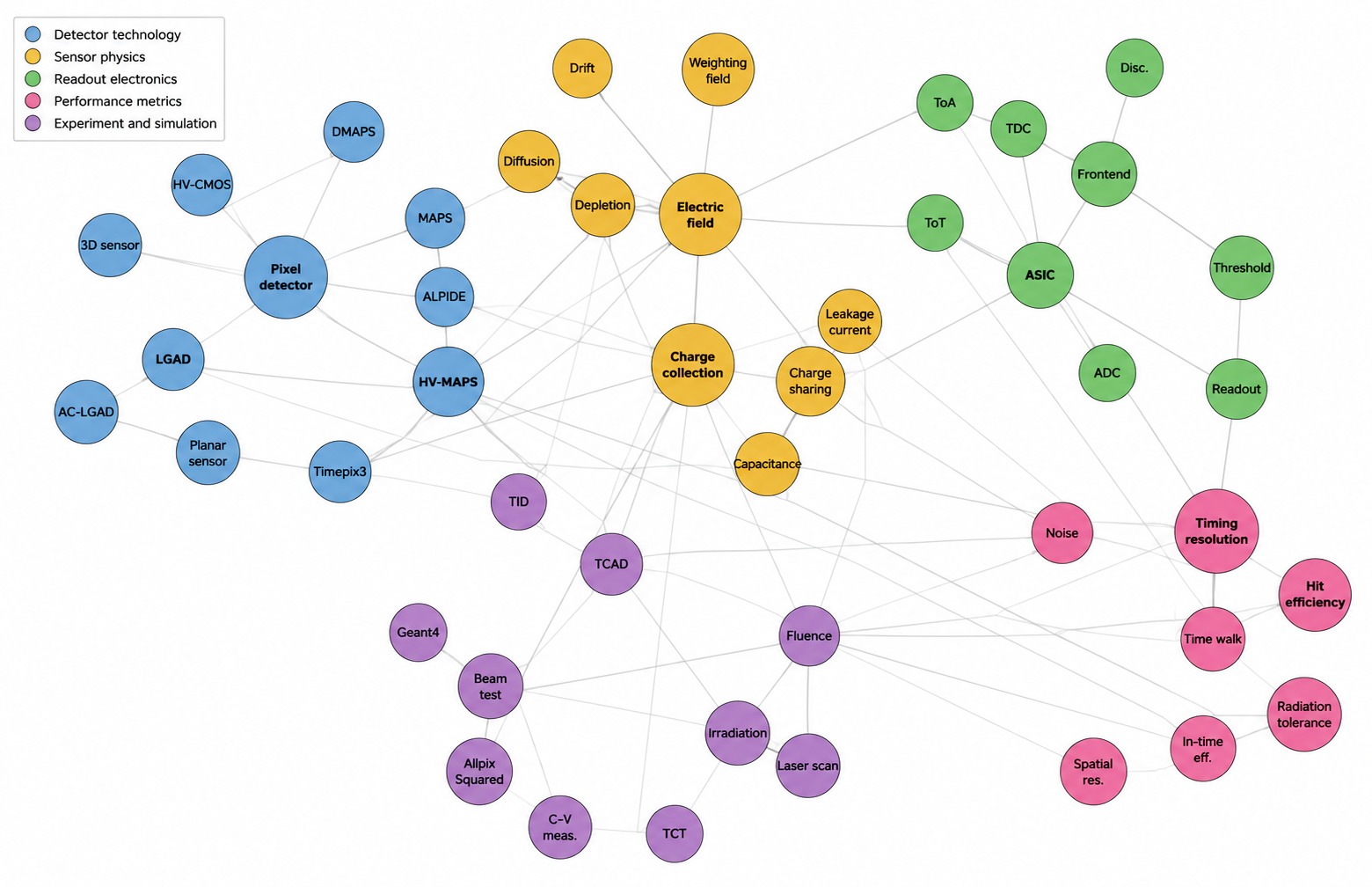}
\caption{Representative detector-entity subgraph used for graph-guided literature exploration. Nodes are grouped into detector technologies, sensor-physics concepts, readout electronics, performance metrics, and experimental workflows. Edges represent literature-derived co-occurrence and metadata relationships used for contextual expansion and detector-concept discovery.}
\label{fig:graph_example}
\end{figure*}

However, semantic proximity is not necessarily equivalent to evidential relevance. Closely related detector technologies may differ substantially in operating conditions, irradiation history, sensor geometry, readout architecture, or measurement methodology. Consequently, unrestricted graph expansion can introduce passages that are topically related but do not constitute the exact supporting evidence required to answer a detector-R\&D question. Figure~\ref{fig:graph_example} shows a representative detector-entity subgraph extracted from the literature corpus. Nodes correspond to detector technologies, sensor-physics concepts, readout electronics, performance metrics, and experimental workflows, while edges represent co-occurrence and metadata relationships used during graph-guided exploration. For this reason, graph-guided exploration is not treated as a replacement for precision-oriented evidence retrieval. Instead, it functions as a controlled contextual-expansion layer that complements the hybrid retrieval backbone while preserving the prioritisation of high-confidence evidence passages.

% However, semantic proximity is not necessarily equivalent to evidential relevance. Closely related detector technologies may differ substantially in operating conditions, irradiation history, sensor geometry, readout architecture, or measurement methodology. Consequently, unrestricted graph expansion can introduce passages that are topically related but do not constitute the exact supporting evidence required to answer a detector-R\&D question. For this reason, graph-guided exploration is not treated as a replacement for precision-oriented evidence retrieval. Instead, it functions as a controlled contextual-expansion layer that complements the hybrid retrieval backbone while preserving the prioritisation of high-confidence evidence passages.
% Figure~\ref{fig:graph_example} shows a representative subgraph extracted from the detector-entity graph. Nodes correspond to detector technologies, sensor-physics concepts, readout electronics, performance metrics, and experimental workflows, while edges represent co-occurrence and metadata relationships extracted from the literature corpus.

\subsection{Evidence-grounded response generation}
\label{sec:response}

Although the primary focus of this work is evidence retrieval, retrieved passages can be supplied to a language model to support interactive literature exploration and evidence-grounded response generation. The language model is not treated as the primary object of evaluation in this study; rather, it serves as a downstream consumer of the retrieved evidence. This design reflects the central objective of the framework, namely reliable evidence access and source traceability for detector-domain literature.

Retrieved passages are provided together with source identifiers and passage metadata, allowing generated responses to remain linked to the supporting detector literature. The language model is instructed to answer only from the retrieved evidence, preserve citation traceability, and abstain when the retrieved material is insufficient. Such behaviour is particularly important for detector-domain questions involving irradiation studies, timing measurements, charge collection, or detector-performance comparisons, where unsupported synthesis across incompatible detector configurations can misrepresent published results.

The prompt template includes the retrieved passages, source identifiers, and an explicit abstention instruction. When no retrieved passage directly supports a requested detector claim under the stated operating conditions, the model is instructed to return a designated abstention response rather than generate unsupported content. Retrieved passages are packed in descending retrieval-rank order up to a fixed context limit, ensuring that the highest-confidence evidence is presented first. Citation traceability is preserved by requiring generated responses to reference the supporting passage identifiers. The complete prompt template is provided in Appendix~\ref{app:prompts}.

\section{Experiments}
\label{sec:experiments}

\subsection{Experimental setup}\label{sec:Experimental setup}

The detector-literature corpus contains 378 source documents covering silicon detector instrumentation, detector simulation, and detector-performance studies. These documents were obtained through automated keyword queries to the inspireHEP database, using broad detector terms such as ``pixel'' and ``silicon detector'', among others. The keywords were kept deliberately permissive to admit as wide a range of detector-instrumentation literature as possible. The queries returned matching publications ordered from most recent to oldest, and the corpus was formed by taking documents in this order up to a fixed size set by the corpus-processing budget rather than by any content criterion. Because the selection depends only on publication date, over which we have no control, and applies no manual inclusion or exclusion of individual documents, it introduces no deliberate bias toward particular detector technologies or results. The benchmark queries and gold annotations were defined afterwards over this fixed corpus, so that answerability reflects genuine corpus coverage rather than documents chosen to match particular queries. After document processing and chunking, the corpus comprises 8,442 indexed passage-level chunks. The corpus spans a broad range of silicon detector technologies, including HV-CMOS and DMAPS devices~\cite{snoeys2014maps,pernegger2017hvcmos}, LGAD and AC-LGAD timing detectors~\cite{pellegrini2014lgad}, ALPIDE and MAPS sensors~\cite{mager2016alpide}, Timepix3-based detectors~\cite{poikela2014timepix3}, and detector-simulation frameworks such as Allpix Squared and Geant4~\cite{spannagel2018allpix,agostinelli2003geant4}.

Evaluation is performed using two complementary benchmark sets designed to assess strict chunk-level evidence retrieval, source-level retrieval, semantic evidence coverage, and abstention behaviour on unsupported detector-domain queries. The benchmark composition is summarised in Table~\ref{tab:benchmark_comp}. The core benchmark contains 60 detector-domain queries, including 48 answerable queries with strict gold evidence and 12 negative queries. The curated extension benchmark contains 56 queries, including 41 answerable queries with strict gold evidence and 15 negative queries. Together, the combined benchmark contains 116 detector-domain queries spanning detector-technology comparison, irradiation studies, timing performance, charge collection, detector simulation, readout architectures, and detector-performance analysis.

\begin{table}[t]
\centering
\caption{Benchmark composition.}
\label{tab:benchmark_comp}
\small
\begin{tabular}{lccc}
\toprule
Benchmark & Total & \shortstack{Answerable\\with strict gold} & Negative \\
\midrule
Core & 60 & 48 & 12 \\
Curated extension & 56 & 41 & 15 \\
Combined & 116 & 89 & 27 \\
\bottomrule
\end{tabular}
\end{table}

Queries are additionally grouped into four reasoning levels covering direct factual retrieval, single-paper parameter retrieval, multi-evidence synthesis, and detector-physics reasoning. This enables evaluation of retrieval performance as a function of reasoning complexity, rather than solely by detector topic. The overall corpus, graph, and benchmark statistics used throughout the retrieval study are summarised in Table~\ref{tab:system_stats}.

\begin{table}[t]
\centering
\caption{System artefacts used throughout the retrieval evaluation.}
\label{tab:system_stats}
\small
\begin{tabular}{lr}
\toprule
Artefact & Value \\
\midrule
Source documents & 378 \\
Text chunks in index & 8442 \\
Entity graph nodes & 2914 \\
Entity graph edges & 18616 \\
Retrieval configurations & 6 \\
Core strict-gold queries & 48 \\
Extension strict-gold queries & 41 \\
Negative/abstention queries total & 27 \\
\bottomrule
\end{tabular}
\end{table}

Strict chunk-level gold labels are manually assigned to passages that directly support the corresponding detector-domain claim under matching operating conditions. Source-level labels test whether the correct publication is retrieved even when the exact supporting passage is missed. Semantic soft-gold annotations provide a diagnostic for topically related but non-exact evidence, while negative queries evaluate whether the system correctly abstains when the corpus does not support a requested detector claim.

To assess the reliability of these strict labels, we performed an additional second-annotation check on a stratified sample of 15 answerable core-benchmark queries, corresponding to 300 query--chunk pairs drawn from the pooled candidate evidence set. The second pass used the same strict criterion: a chunk was labelled as gold only when it directly supported the detector-domain claim under the relevant operating conditions, and not when it was merely topically related. After calibration of this relevance criterion, agreement with the primary labels reached Cohen's $\kappa=0.619$, with raw agreement of 0.897, positive-specific agreement of 0.680, and negative-specific agreement of 0.938. Because gold passages are sparse within each candidate pool, the positive- and negative-specific agreement values are reported alongside $\kappa$.

Retrieval quality is evaluated using strict Hit@5~\cite{manning2008ir} and Mean Reciprocal Rank (MRR)~\cite{Voorhees1999} as the primary metrics. Hit@5 measures whether the annotated gold evidence chunk appears within the top five retrieved results, while MRR quantifies the ranking position of the first relevant passage. To distinguish exact evidence retrieval from broader literature discovery, we additionally report Paper@5 and Soft@5. Paper@5 measures whether the correct source publication is retrieved, while Soft@5 measures retrieval of semantically relevant evidence that may not exactly match the annotated gold passage. For negative queries, abstention accuracy is reported as the fraction of unsupported queries on which the system correctly abstains.

\subsection{Retrieval performance evaluation}
\label{sec:retrieval_eval}

We first optimise the hybrid retrieval configuration and then compare sparse, dense, hybrid and graph-guided retrieval strategies across the detector-domain benchmarks. The objective is to determine whether graph-enhanced retrieval improves strict evidence ranking, or whether its primary value lies in literature exploration and detector-concept navigation.

The default fusion weights $w_d=0.9$ and $w_b=0.1$ in Eq.~\eqref{eq:rrf} are selected through a grid search over dense--sparse weight combinations on the answerable benchmark queries. Increasing the dense contribution generally improves ranking quality, while a small lexical component remains beneficial for preserving sensitivity to specialised detector terminology and exact detector names. The configuration $w_d=0.9$ and $w_b=0.1$ achieves the best overall performance and is therefore adopted throughout the remainder of the study.

\begin{figure*}[t]
\centering
\includegraphics[width=\textwidth]{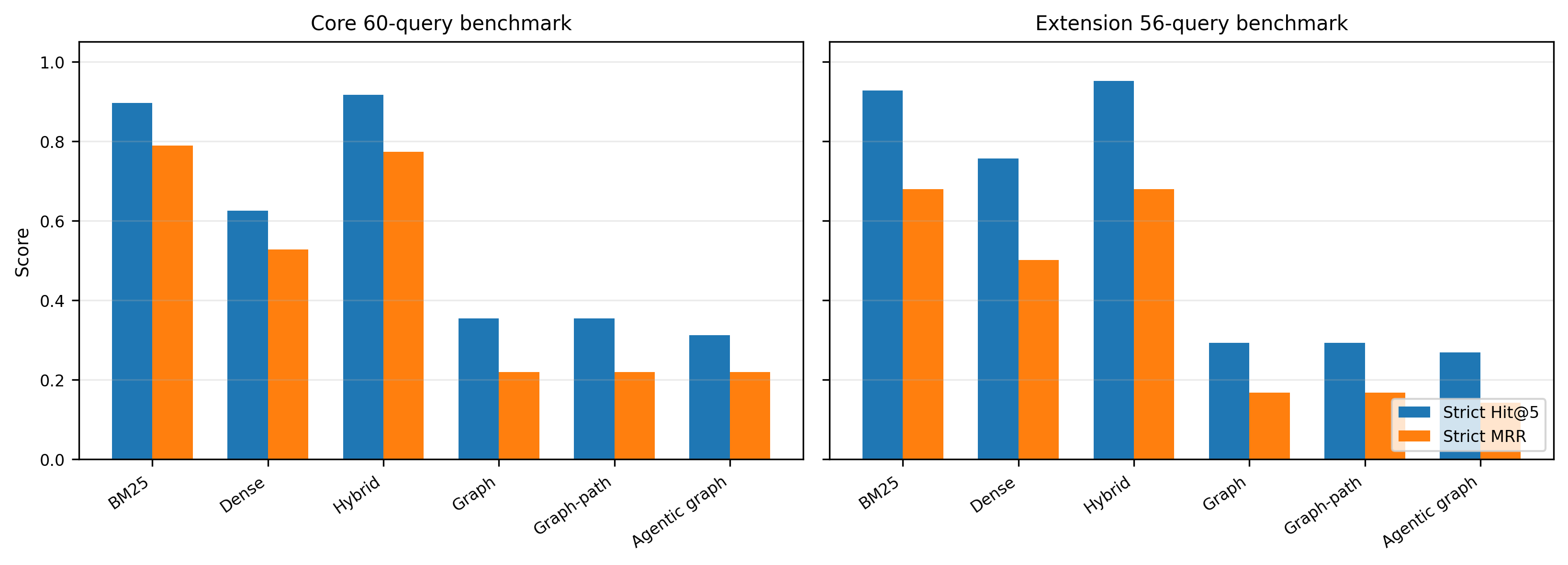}
\caption{Strict Hit@5 and MRR for the core and extension benchmarks. Hybrid retrieval provides the strongest strict chunk-level evidence retrieval, while BM25 remains highly competitive because of exact detector terminology.}
\label{fig:main_results}
\end{figure*}

\begin{figure*}[t]
\centering
\includegraphics[width=\textwidth]{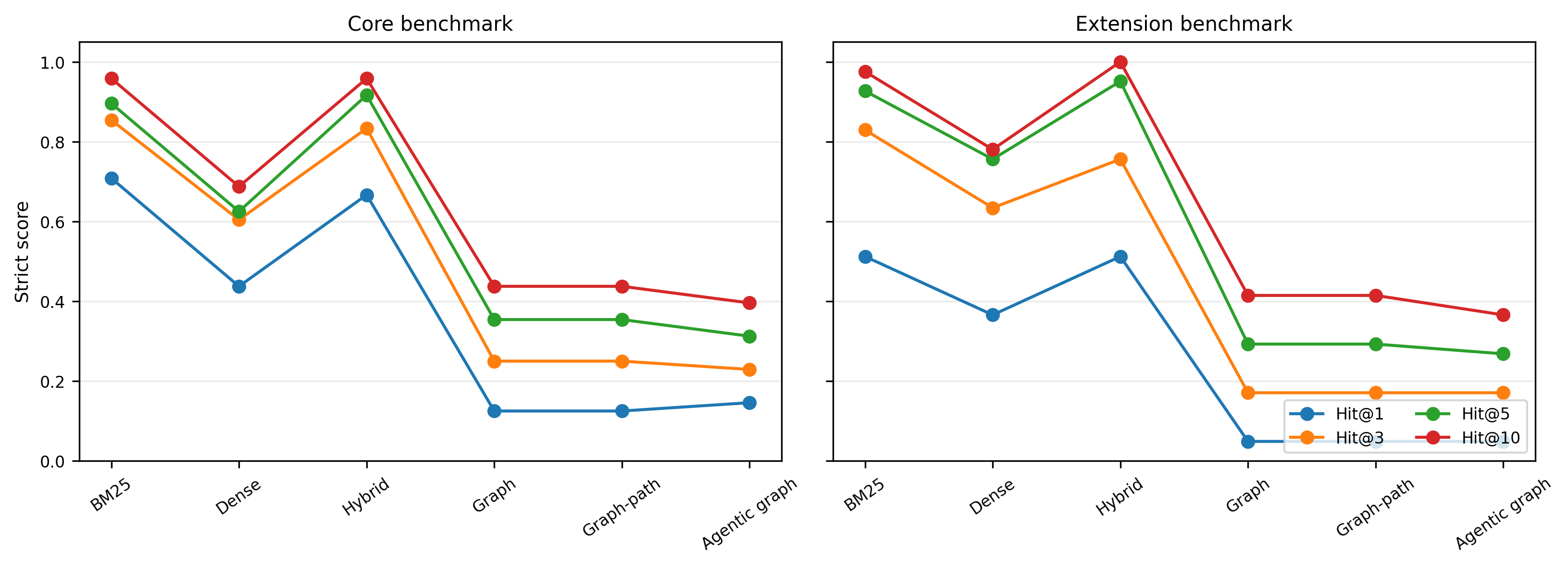}
\caption{Strict Hit@k profiles for the core and extension benchmarks. The hybrid method consistently reaches the highest or joint-highest recall at larger $k$, while graph-heavy configurations underperform on strict chunk matching across all $k$.}
\label{fig:hitk}
\end{figure*}

We first compare all retrieval configurations under strict chunk-level evaluation and then examine their broader source-level and semantic retrieval behaviour. Table~\ref{tab:main_results} presents the main retrieval results for all six configurations on the core and extension benchmarks. Figures~\ref{fig:main_results} and~\ref{fig:hitk} show the strict Hit@5, MRR and Hit@k behaviour.

\begin{table*}[t]
\centering
\caption{Main retrieval results on the 60-query core benchmark and 56-query curated extension benchmark. Strict metrics use chunk-level gold evidence. Paper and soft metrics are reported as complementary coverage diagnostics. The best score in each column is shown in bold.}
\label{tab:main_results}
\scriptsize
\setlength{\tabcolsep}{5.5pt}
\begin{tabular}{@{}lcccccccc@{}}
\toprule
 & \multicolumn{4}{c}{Core 60} & \multicolumn{4}{c}{Extension 56}\\
\cmidrule(lr){2-5}\cmidrule(lr){6-9}
Method & Hit@5 & MRR & Paper@5 & Soft@5 & Hit@5 & MRR & Paper@5 & Soft@5\\
\midrule
BM25 & 0.896 & \textbf{0.789} & \textbf{0.983} & \textbf{0.933} & 0.927 & \textbf{0.679} & \textbf{0.946} & \textbf{0.857}\\
Dense & 0.625 & 0.527 & 0.917 & 0.833 & 0.756 & 0.501 & 0.839 & 0.804\\
Hybrid & \textbf{0.917} & 0.773 & 0.967 & 0.917 & \textbf{0.951} & \textbf{0.679} & \textbf{0.946} & \textbf{0.857}\\
Graph & 0.354 & 0.220 & 0.767 & 0.733 & 0.293 & 0.167 & 0.554 & 0.518\\
Graph-path & 0.354 & 0.220 & 0.767 & 0.733 & 0.293 & 0.167 & 0.554 & 0.518\\
Agentic graph & 0.312 & 0.220 & 0.767 & 0.733 & 0.268 & 0.142 & 0.589 & 0.536\\
\bottomrule
\end{tabular}
\end{table*}
\begin{table*}[t]
\centering
\caption{Bootstrap 95\% confidence intervals for strict retrieval performance on the
48 answerable queries in the core benchmark, estimated using 10,000 bootstrap
resamples.}
\label{tab:bootstrap_ci}
\begin{tabular}{lcccc}
\toprule
Method & Hit@5 & 95\% CI & MRR & 95\% CI \\
\midrule
BM25 & 0.896 & [0.792, 0.979] & 0.789 & [0.686, 0.880] \\
Dense & 0.625 & [0.479, 0.750] & 0.527 & [0.399, 0.651] \\
Hybrid & 0.917 & [0.833, 0.979] & 0.773 & [0.672, 0.867] \\
Graph & 0.354 & [0.229, 0.500] & 0.220 & [0.130, 0.321] \\
Graph-path & 0.354 & [0.229, 0.500] & 0.220 & [0.129, 0.318] \\
Agentic graph & 0.312 & [0.188, 0.438] & 0.220 & [0.126, 0.322] \\
\bottomrule
\end{tabular}
\end{table*}

To assess statistical robustness, Table~\ref{tab:bootstrap_ci} reports 95\% bootstrap confidence intervals for the strict retrieval metrics using 10,000 bootstrap resamples. The resulting intervals confirm that hybrid retrieval consistently achieves the strongest strict evidence-ranking performance, while graph-based retrieval methods remain substantially below the lexical and hybrid baselines.

\begin{table}[t]
\centering
\caption{Paired Wilcoxon signed-rank tests comparing hybrid retrieval against
alternative retrieval methods using query-level strict Hit@5 results on the
answerable benchmark queries.}
\label{tab:significance}
\begin{tabular}{lc}
\toprule
Comparison & $p$-value \\
\midrule
Hybrid vs BM25 & 0.788 \\
Hybrid vs Dense & $3.24\times10^{-3}$ \\
Hybrid vs Graph & $5.52\times10^{-7}$ \\
Hybrid vs Graph-path & $5.52\times10^{-7}$ \\
Hybrid vs Agentic graph & $1.62\times10^{-7}$ \\
\bottomrule
\end{tabular}
\end{table}

The statistical comparison in Table~\ref{tab:significance} shows that hybrid retrieval significantly outperforms dense and graph-based retrieval methods, while the difference between hybrid retrieval and BM25 is not statistically significant. This result reflects the strong performance of lexical retrieval in detector-domain literature, where detector names, acronyms and operating-condition terminology remain highly informative.

Hybrid retrieval provides the strongest strict chunk-level evidence ranking across both benchmarks, reaching Hit@5 of 0.917 on the core benchmark and 0.951 on the extension benchmark. BM25 remains highly competitive, reflecting the acronym-rich and terminology-stable nature of detector literature. Dense retrieval performs worse under strict chunk-level evaluation, although its Paper@5 scores show that it frequently retrieves the correct publication while missing the exact supporting passage. Graph, graph-path and agentic graph configurations substantially underperform the lexical and hybrid baselines under strict evaluation. The agentic graph configuration performs slightly worse than the graph-path
baseline, indicating that additional query decomposition does not compensate for the loss of ranking precision introduced by graph expansion. Their Paper@5 and Soft@5 scores remain considerably higher than their strict Hit@5 values, indicating that graph expansion often retrieves semantically related passages or correct source papers without ranking the exact gold chunk highly enough. This distinction between semantic exploration and exact evidence retrieval is a recurring feature of the detector-domain corpus.

\begin{figure}[t]
\centering
\includegraphics[width=0.90\linewidth]{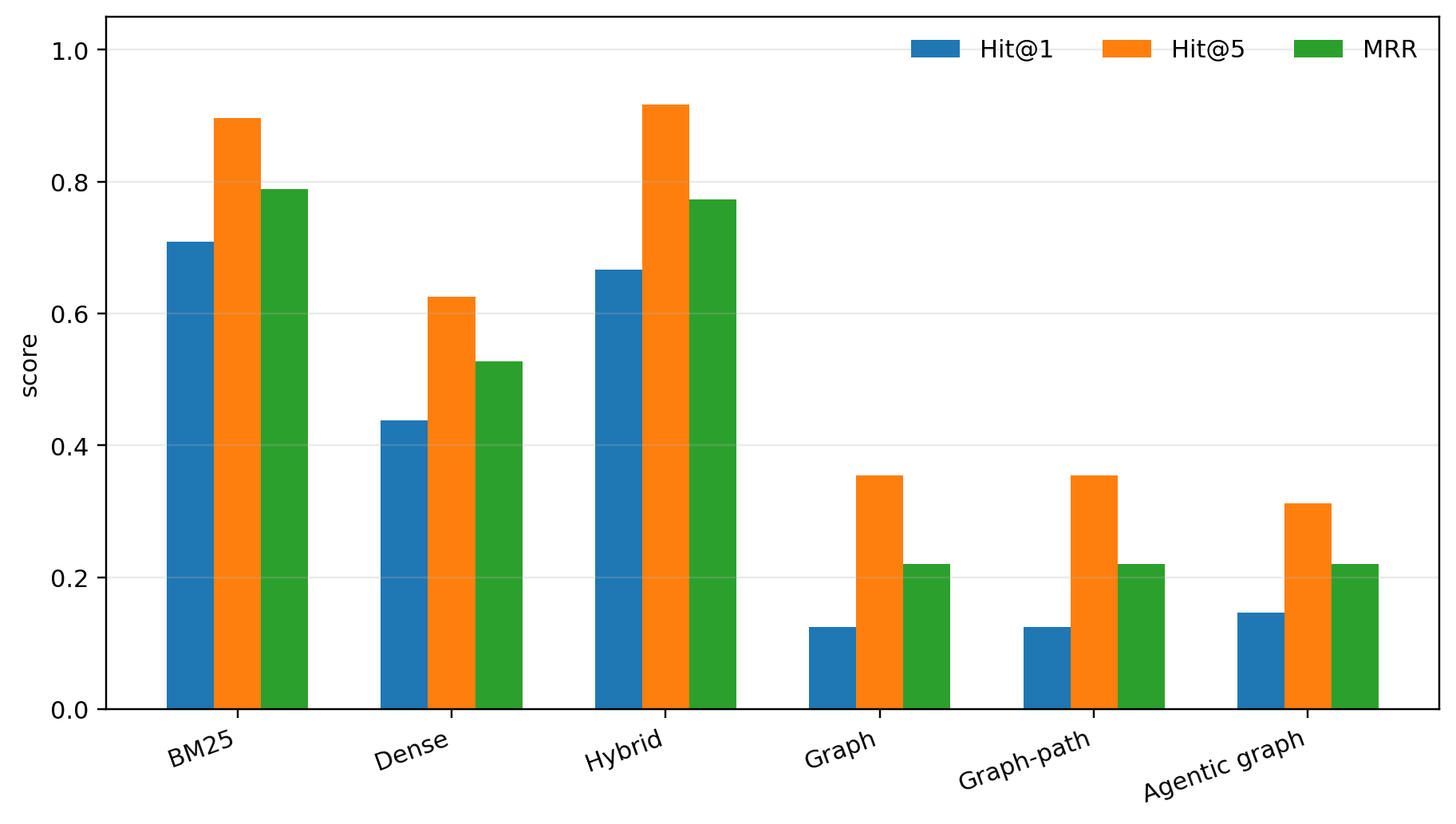}
\caption{Ablation of hybrid retrieval components. Removing BM25 or dense retrieval reduces strict Hit@5. The full hybrid configuration provides the strongest performance, with BM25 contributing exact terminology matching and dense retrieval contributing complementary semantic coverage.}
\label{fig:ablation}
\end{figure}

Figure~\ref{fig:ablation} confirms that the sparse and dense retrieval components are complementary. BM25 provides the dominant strict-matching signal, while dense retrieval improves coverage for paraphrased detector-physics queries and semantically related detector concepts. The full hybrid configuration therefore provides the most robust retrieval performance across the benchmark.

\begin{figure*}[t]
\centering
\includegraphics[width=0.90\textwidth]{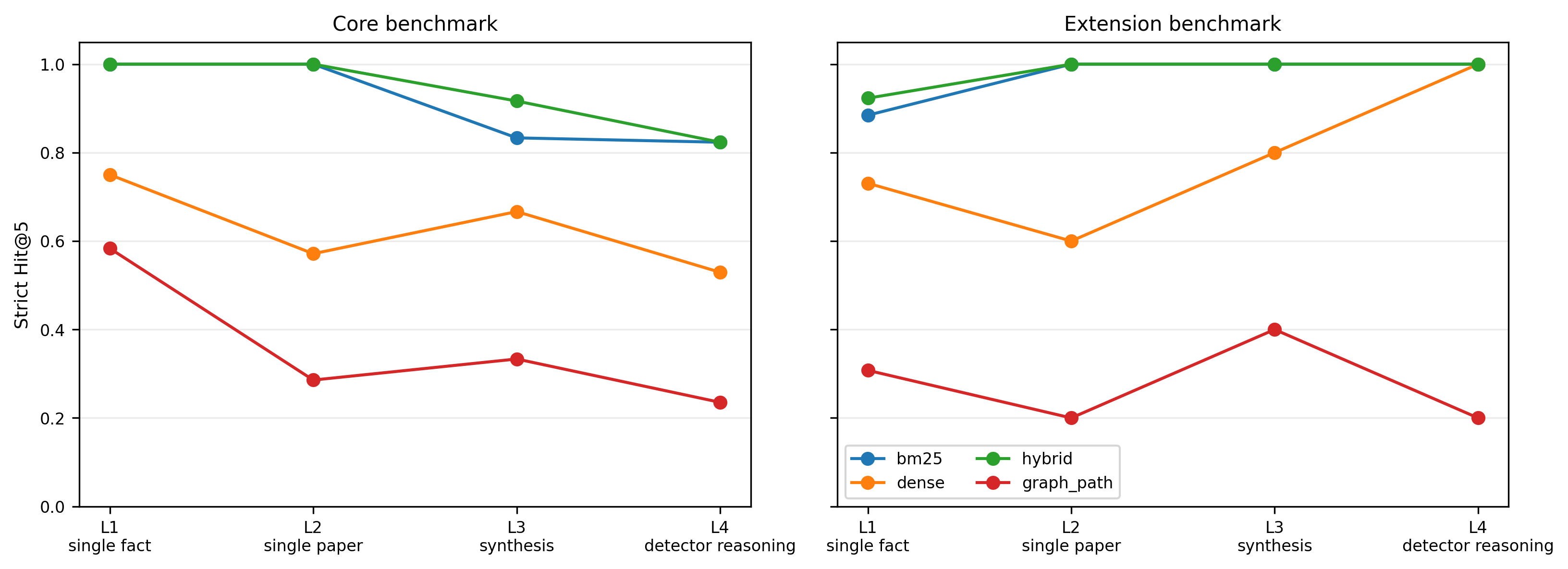}
\caption{Retrieval performance as a function of reasoning complexity. Performance generally decreases as queries require increasingly complex evidence synthesis, while hybrid retrieval remains the most stable configuration across all reasoning levels.}
\label{fig:reasoning_level}
\end{figure*}

Figure~\ref{fig:reasoning_level} shows retrieval performance as a function of reasoning complexity. All retrieval configurations perform strongly on direct factual retrieval and single-paper parameter queries, while performance decreases for multi-evidence synthesis and detector-physics reasoning tasks. Hybrid retrieval remains the most stable configuration across all reasoning levels, indicating that combining lexical and semantic retrieval signals improves robustness as query complexity increases.
\begin{figure}[t]
\centering
\includegraphics[width=0.92\linewidth]{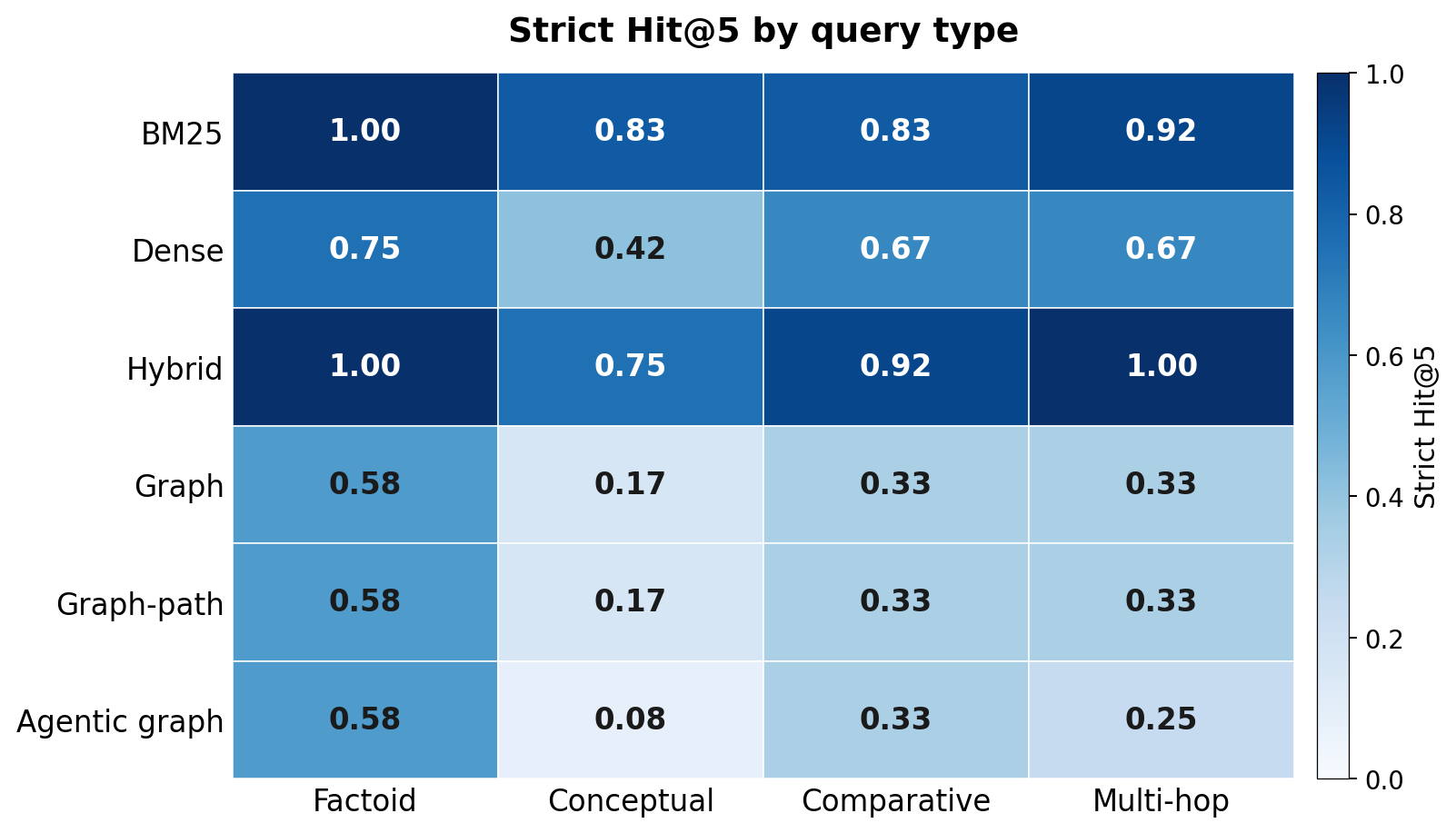}
\caption{Retrieval performance across detector-domain query categories. Different detector topics exhibit different retrieval characteristics, reflecting variations in terminology stability, evidence density, and conceptual complexity.}
\label{fig:query_heatmap}
\end{figure}

Figure~\ref{fig:query_heatmap} provides a complementary view by grouping queries according to detector topic rather than reasoning complexity. Queries involving detector technologies, irradiation studies, and timing measurements generally achieve the strongest retrieval performance, reflecting the stable terminology and well-defined evidence structure of these topics. Lower performance is observed for broader detector-physics and cross-technology comparison queries, where supporting evidence is often distributed across multiple publications.

\begin{figure}[t]
\centering
\includegraphics[width=0.90\linewidth]{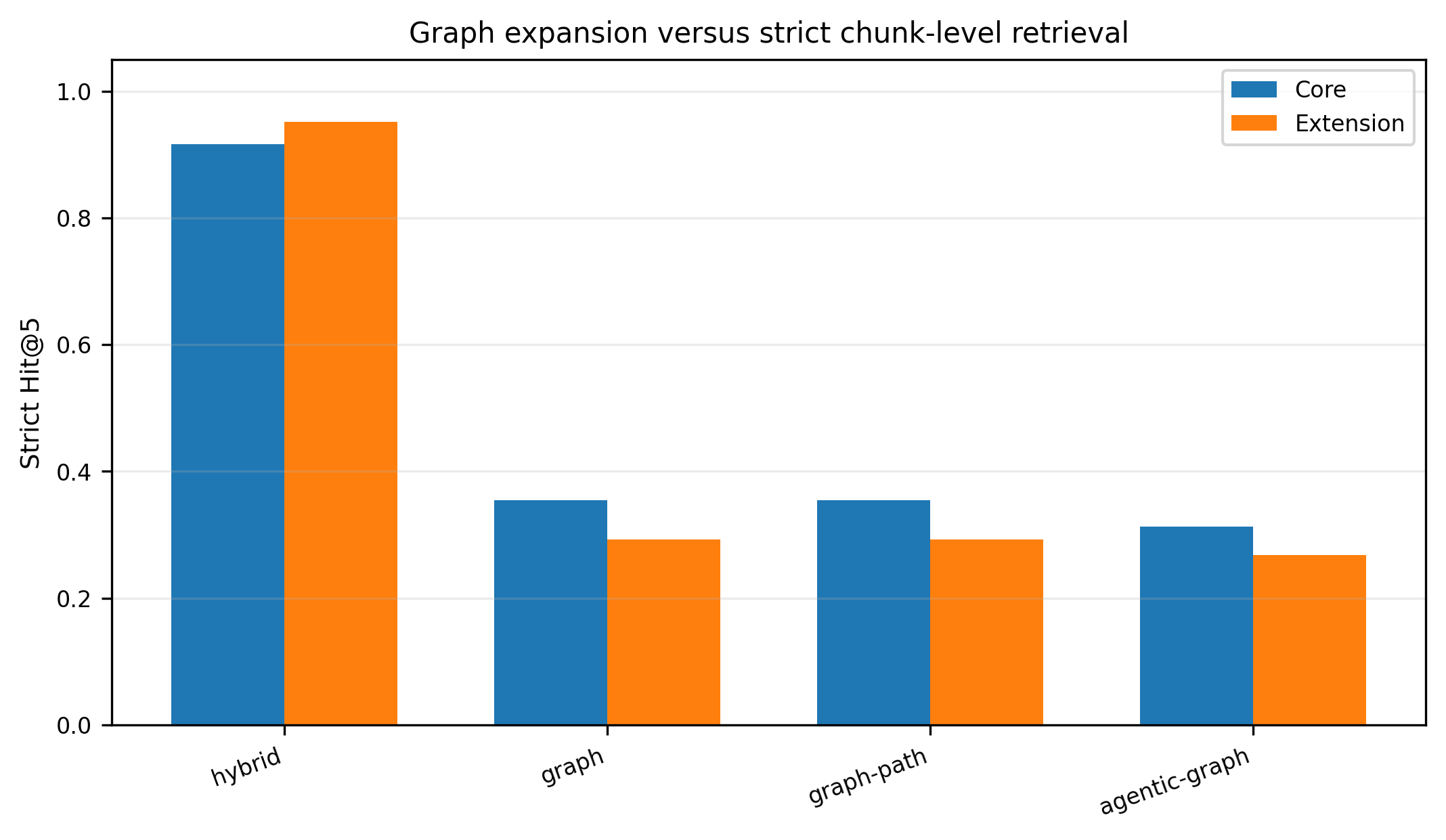}
\caption{Effect of graph expansion on strict retrieval performance. While graph traversal increases semantic coverage, excessive expansion dilutes exact evidence ranking and reduces strict chunk-level retrieval accuracy.}
\label{fig:graph_dilution}
\end{figure}

\begin{figure*}[t]
\centering
\includegraphics[width=0.88\textwidth]{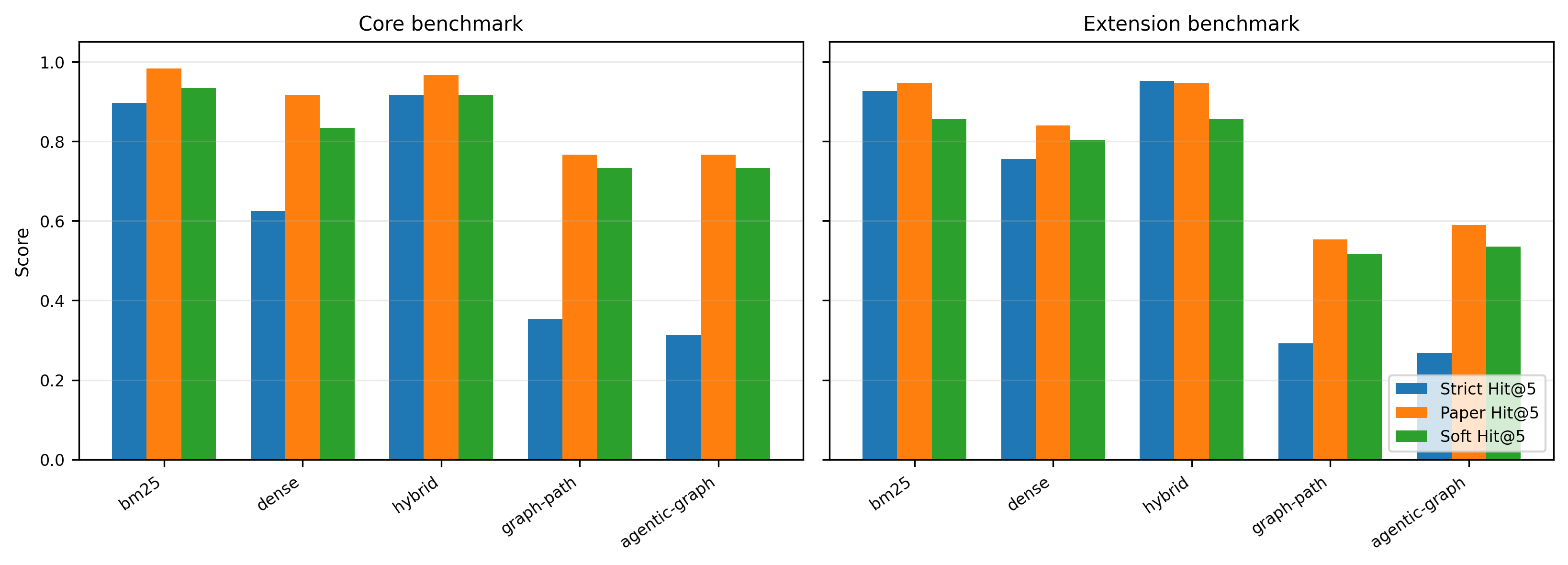}
\caption{Comparison of strict chunk-level retrieval, source-level retrieval, and semantic soft-gold retrieval. Graph-based retrieval recovers relevant publications and semantically related passages more effectively than exact supporting evidence.}
\label{fig:strict_paper_soft}
\end{figure*}

Graph-based retrieval exhibits a markedly different behaviour from the lexical and hybrid baselines. While strict chunk-level retrieval performance is substantially lower, graph-guided methods achieve comparatively stronger source-level and semantic coverage. Figure~\ref{fig:graph_dilution} illustrates this behaviour. As graph expansion becomes more aggressive, semantically related detector concepts are increasingly retrieved, but exact supporting evidence is displaced by neighbouring passages. In practice, graph expansion frequently identifies publications and passages that are relevant to the broader detector concept, even when the exact supporting evidence is not ranked highly enough to satisfy strict evaluation criteria. This behaviour suggests that graph retrieval is better suited to literature exploration and detector-concept navigation than to exact evidence ranking. Figure~\ref{fig:strict_paper_soft} further illustrates the distinction between strict evidence retrieval and broader literature discovery. While graph-based methods perform poorly under strict chunk-level evaluation, their source-level and semantic retrieval scores remain substantially higher. This indicates that graph expansion frequently reaches the correct publication or a semantically related passage, but does not reliably rank the exact supporting evidence highly enough for strict retrieval tasks. The result reinforces the view that graph retrieval functions most effectively as an exploratory literature-navigation tool rather than a primary evidence ranker.

\begin{table}[t]
\centering
\caption{Failure taxonomy for hybrid retrieval on the 48 answerable core benchmark
queries.}
\label{tab:failure_taxonomy}
\begin{tabular}{lcc}
\toprule
Failure category & Count & Fraction \\
\midrule
Conceptual synthesis queries & 3 & 75\% \\
Cross-technology comparison & 1 & 25\% \\
Detector-condition mismatch & 0 & 0\% \\
Acronym ambiguity & 0 & 0\% \\
\midrule
Total failures & 4 & 100\% \\
\bottomrule
\end{tabular}
\end{table}
\paragraph{Failure analysis.}

Table~\ref{tab:failure_taxonomy} summarises the dominant failure modes observed for the hybrid retriever. Most remaining errors arise from conceptual synthesis and cross-technology comparison queries rather than terminology matching. In these cases, the required evidence is distributed across multiple publications and cannot always be recovered through retrieval of a single supporting passage. Terminology ambiguity contributes comparatively little to the residual error rate, indicating that lexical and hybrid retrieval largely resolves detector-specific vocabulary matching. The principal remaining limitation therefore lies in higher-level scientific reasoning rather than first-stage evidence retrieval. These observations suggest that future improvements are more likely to arise from multi-passage evidence synthesis and reasoning-aware retrieval strategies than from further optimisation of lexical matching.
\begin{figure*}[t]
\centering
\includegraphics[width=\textwidth]{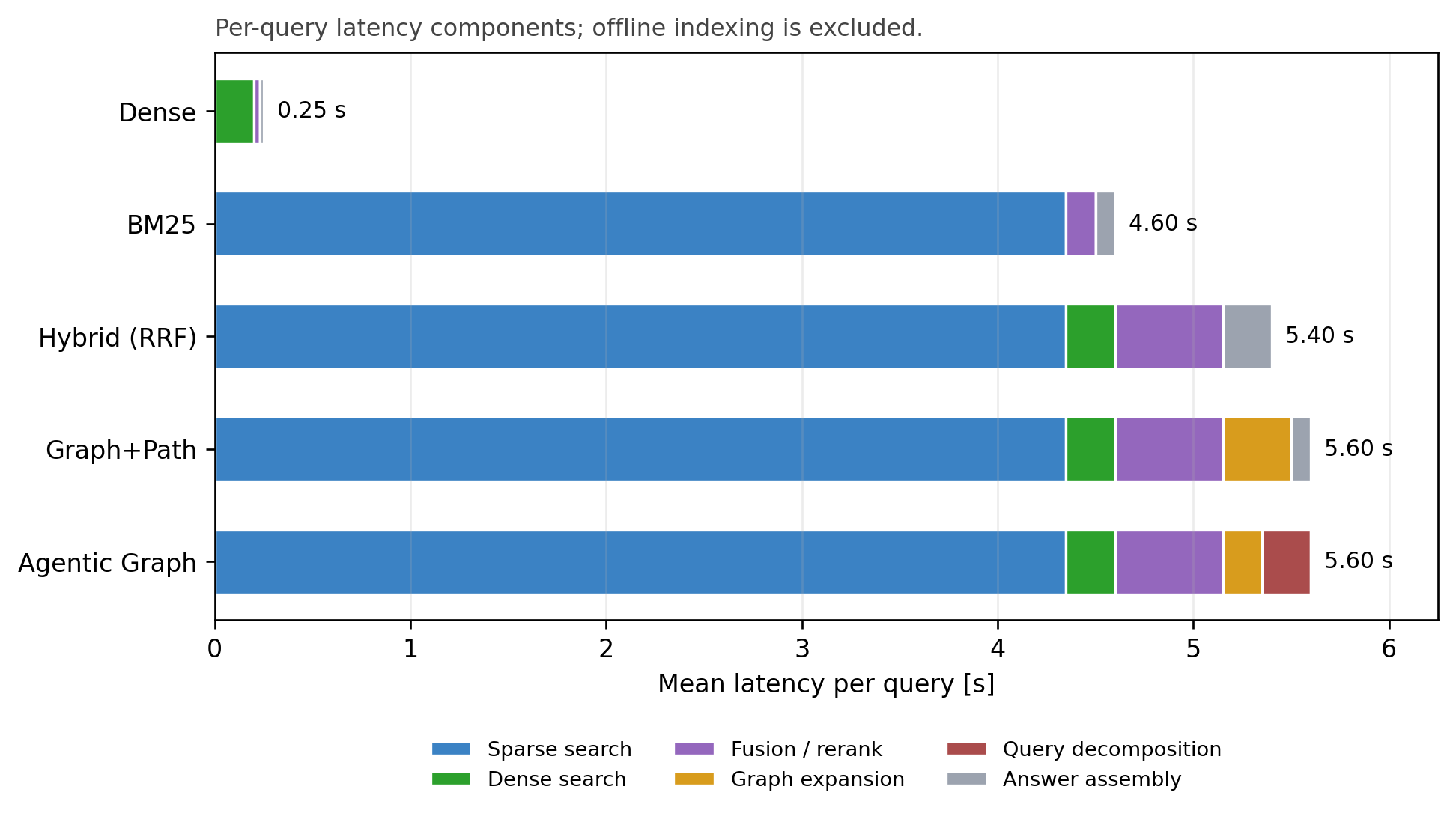}
\caption{Per-query latency decomposition by pipeline stage. Sparse and dense search dominate the retrieval cost, while graph configurations add further graph-expansion and query-decomposition stages. The breakdown shows where the additional latency of graph-based retrieval originates rather than its total cost alone.}
\label{fig:latency}
\end{figure*}
\begin{figure*}[t]
\centering
\includegraphics[width=\textwidth]{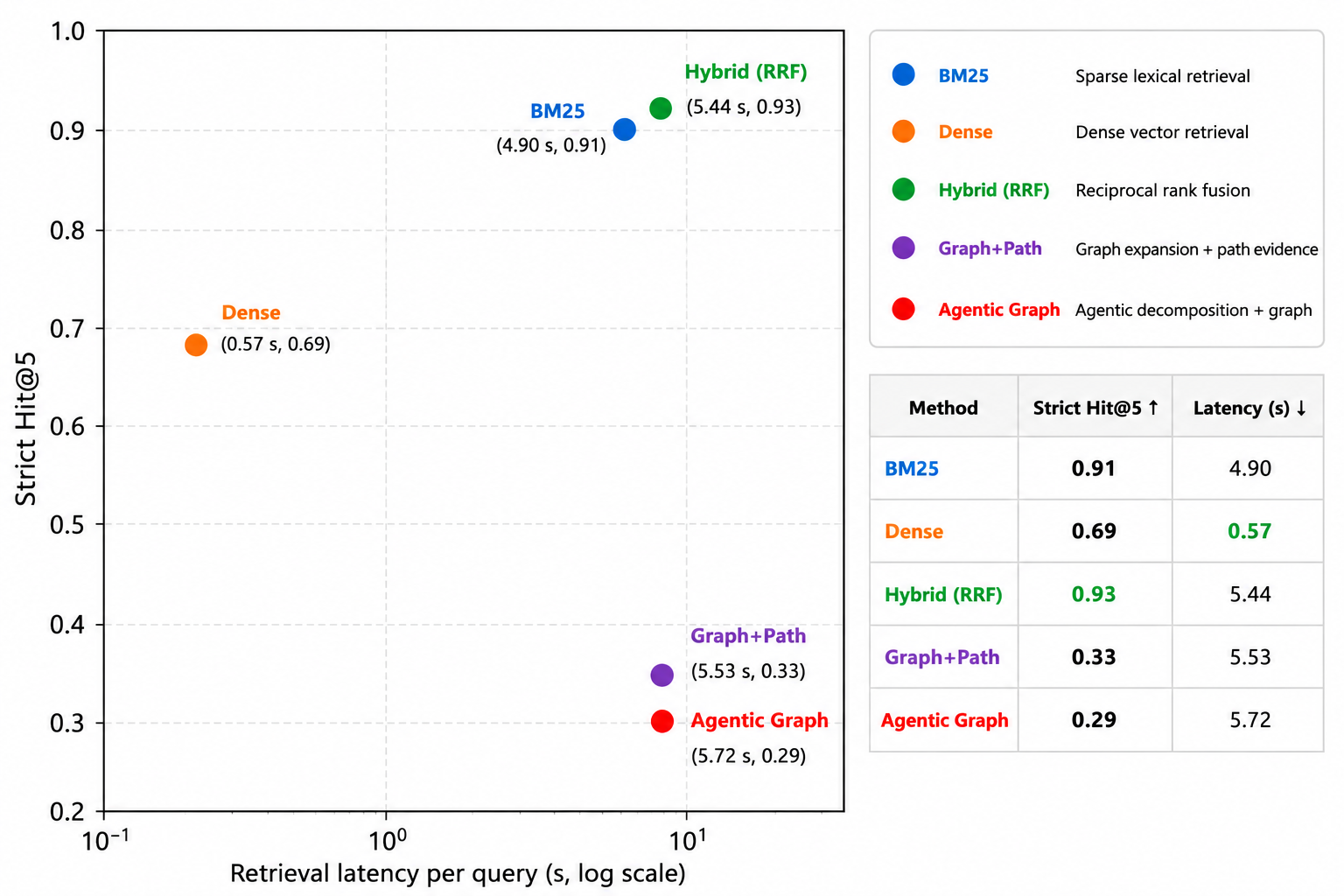}
\caption{Trade-off between strict retrieval performance and query latency. Hybrid retrieval achieves the strongest strict Hit@5 while introducing only a modest latency increase relative to BM25. Graph-based retrieval configurations incur additional latency without improving strict evidence-ranking performance.}
\label{fig:tradeoff}
\end{figure*}
\paragraph{Retrieval-performance trade-off.}

Figure~\ref{fig:tradeoff} summarises the trade-off between retrieval effectiveness and query latency. Dense retrieval provides the lowest latency but substantially weaker strict evidence retrieval. BM25 achieves strong retrieval performance with moderate latency, reflecting the stability of detector terminology and acronyms. Hybrid retrieval occupies the most favourable operating point, achieving the highest strict Hit@5 while introducing only a modest latency increase relative to BM25. In contrast, graph-based retrieval configurations require additional graph-expansion and query-processing stages but do not improve strict retrieval performance. These results further support hybrid retrieval as the preferred default configuration for evidence-grounded detector-literature access.
\paragraph{Latency decomposition.}

Figure~\ref{fig:latency} decomposes per-query latency by pipeline stage. Sparse and dense retrieval account for most of the cost in the BM25, dense, and hybrid configurations, so hybrid retrieval inherits only a small fusion overhead on top of its two component retrievers. The graph and agentic-graph configurations add distinct graph-expansion and query-decomposition stages on top of the same retrieval backbone, which is the origin of their higher latency seen in Figure~\ref{fig:tradeoff}. This decomposition clarifies that the additional cost of graph-based retrieval comes from contextual expansion rather than from the core evidence-retrieval step.
\subsection{Abstention behaviour on negative queries}
\label{sec:abstention}

Beyond strict evidence ranking, a trustworthy detector-literature assistant must recognise when the corpus does not support a requested claim and abstain rather than generate unsupported conclusions. The negative queries in each benchmark are therefore constructed so that no gold evidence passage exists in the corpus. The system is expected to return a predefined abstention marker when the retrieved evidence is insufficient to support the queried detector-domain claim, and abstention accuracy is measured as the fraction of unsupported queries on which the system correctly abstains.

On the 12 negative queries of the core benchmark, the hybrid configuration achieves an abstention accuracy of 1.0, correctly declining to answer every unsupported query under the grounded-generation prompt. This indicates that, when retrieval fails to identify supporting evidence, the grounding constraint reliably triggers abstention rather than unsupported synthesis. Such behaviour is particularly important for detector R\&D applications, where fabricated detector-performance claims or unsupported technology comparisons could lead to misleading scientific conclusions.

\begin{figure}[t]
\centering
\includegraphics[width=0.88\linewidth]{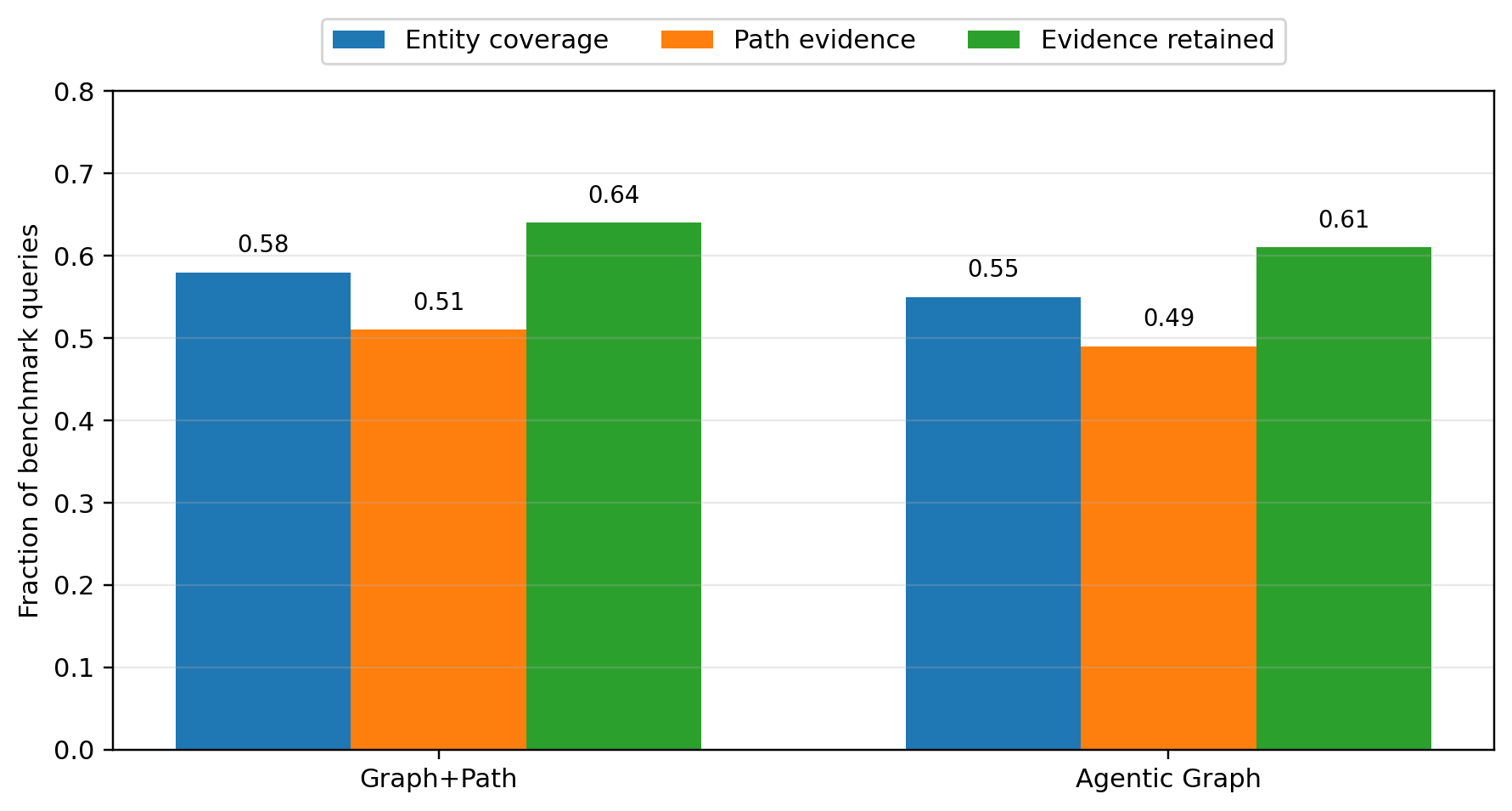}
\caption{Graph coverage and abstention behaviour on negative queries. Graph-based retrieval shows higher false-positive rates than lexical and hybrid retrieval because entity expansion can retrieve plausible but unsupported neighbouring evidence.}
\label{fig:graphabstention}
\end{figure}

Figure~\ref{fig:graphabstention} illustrates the effect of graph expansion on abstention behaviour. Graph-guided retrieval is more likely to surface semantically related passages even when the corpus does not contain evidence supporting the queried claim. While such behaviour can be useful for literature exploration, it weakens the abstention signal required for reliable evidence-grounded retrieval. This observation is consistent with the strict-ranking results and further supports the use of graph expansion as an optional exploratory layer rather than the default evidence ranker.

This study focuses on retrieval quality and abstention behaviour rather than end-to-end answer scoring. In the detector-literature setting, retrieval accuracy is the primary determinant of evidence-grounded behaviour because answer generation is constrained to operate only on retrieved evidence. Evaluating retrieval quality therefore provides a direct assessment of the reliability of the underlying literature-access framework for silicon detector R\&D.
\section{Discussion and Conclusion}
\label{sec:conclusion}

This work presents a grounded evidence-retrieval benchmark and reproducible hybrid retrieval framework for silicon pixel detector literature. The benchmark targets a practical detector-R\&D problem: locating source-attributed, passage-level evidence within a specialised literature corpus characterised by stable detector terminology, configuration-dependent measurements, and detector-physics reasoning. To support this task, we construct a detector-domain benchmark with strict chunk-level annotations, source-level diagnostics, semantic soft-gold evaluation, and negative-query abstention tests, together with a retrieval framework combining BM25 sparse retrieval, dense semantic retrieval, hybrid reciprocal-rank fusion, and graph-guided retrieval strategies.

The principal finding is that hybrid sparse--dense retrieval provides the strongest strict evidence-ranking performance. Hybrid retrieval achieves Hit@5 values of 0.917 on the core benchmark and 0.951 on the extension benchmark, consistently outperforming dense retrieval and all graph-based retrieval configurations under strict chunk-level evaluation. BM25 remains highly competitive because of the lexical stability of detector terminology and acronyms, while dense retrieval improves robustness to paraphrased detector-physics descriptions. In contrast, graph-based retrieval substantially improves semantic exploration and source discovery but does not improve strict passage-level evidence ranking. Similarly, the agentic graph configuration increases system complexity through
query decomposition without providing a measurable gain in strict evidence
retrieval, suggesting that decomposition alone is insufficient when the primary
limitation lies in evidence ranking rather than query understanding. The results therefore reveal a clear distinction between literature exploration and exact evidence retrieval within detector instrumentation literature.

This distinction has practical implications for detector R\&D. For detector instrumentation studies, the most useful retrieval system is not necessarily the one that retrieves the broadest semantic neighbourhood, but the one that reliably identifies the exact supporting evidence under the relevant detector configuration and operating conditions. Detector-performance claims often depend on specific irradiation fluences, bias voltages, temperatures, readout settings, and sensor geometries, making precise evidence attribution essential. This requirement is expected to become increasingly important for future detector-development programmes. Beyond any single experiment, a reproducible and openly released evidence benchmark also lets the wider community track and compare retrieval quality as these tools mature, rather than relying on bespoke or unpublished evaluations.

Several limitations remain. The benchmark labels were produced through a primary annotation pass and do not yet include full multi-annotator adjudication across all query--chunk pairs. However, the calibrated second-annotation check described in Section~\ref{sec:experiments} gives Cohen's $\kappa=0.619$ on 300 sampled pairs, providing a quantitative reliability check for the strict relevance criterion. The reported metrics should therefore still be interpreted under the current annotation scheme and may shift under full reconciliation. The negative queries were constructed to be unsupported by the corpus, but they were not independently exhaustively verified against every chunk. The corpus is a keyword-matched sample ordered by recency and capped by processing budget rather than an exhaustive collection, so coverage of the field is representative but weighted toward recent literature, and it is further restricted to text-extracted evidence from published literature. The dense retriever uses a general-purpose sentence embedding model rather than a detector-domain encoder.

Overall, the results demonstrate that detector-domain retrieval should be
evaluated at the level of supporting evidence rather than only at the level
of documents or generated answers. We release the benchmark, evaluation
framework, and reproducible retrieval pipeline as research-software
artefacts, so that evidence-grounded retrieval for scientific literature can
be measured, audited, and adapted to other detector-literature corpora
rather than assumed. In doing so, the work contributes a reusable component
to the computing and software ecosystem for high-energy physics, and a
practical capability for the detector R\&D and future-collider programmes
whose design decisions increasingly depend on integrating quantitative
evidence across many sensor technologies, irradiation campaigns, beam tests,
and simulation studies.

\acknowledgments
M. Kenzie and T. Gao are supported by UK Research and Innovation under grant \texttt{\#EP/X014746/2}.

\section*{Data and code availability}
The code, benchmark annotations, and retrieval outputs supporting this study will be made publicly available upon acceptance of the article.

\appendix
\section{Grounding and Abstention Prompt}
\label{app:prompts}

The grounded generation component receives the user query together with the retrieved evidence passages and associated source metadata. The model is instructed to answer only from the supplied evidence, preserve source traceability, and abstain when the retrieved material does not support the requested detector-domain claim.

The prompt follows the general structure:

\begin{verbatim}
You are a detector-literature assistant.

Use only the supplied evidence passages.

Requirements:
1. Do not use external knowledge.
2. Cite supporting passages when answering.
3. If the evidence is insufficient,
   return ABSTAIN.
4. Do not infer detector performance
   beyond what is explicitly stated.

Question:
{query}

Retrieved Evidence:
{retrieved_chunks}

Answer:
\end{verbatim}

For negative-query evaluation, the model is expected to return the predefined token

\begin{verbatim}
ABSTAIN
\end{verbatim}

when no retrieved passage directly supports the queried detector-domain claim.

\section{Retrieval Metrics}
\label{app:metrics}

The primary evaluation metrics used throughout this work are strict chunk-level Hit@5 and Mean Reciprocal Rank (MRR).

\paragraph{Hit@5.}

Hit@5 evaluates whether at least one gold evidence chunk appears among the top five retrieved passages:

\begin{equation}
\mathrm{Hit@5}
=
\frac{1}{N}
\sum_{i=1}^{N}
\mathbf{1}
\!\left(
\mathrm{rank}_i \le 5
\right),
\end{equation}

where $N$ is the number of benchmark queries and $\mathrm{rank}_i$ is the rank of the first gold chunk for query $i$.

\paragraph{Mean Reciprocal Rank.}

MRR measures the average reciprocal rank of the first retrieved gold passage:

\begin{equation}
\mathrm{MRR}
=
\frac{1}{N}
\sum_{i=1}^{N}
\frac{1}{\mathrm{rank}_i}.
\end{equation}

\paragraph{Paper@5.}

Paper@5 evaluates whether the correct source publication is retrieved within the top five results, regardless of whether the exact gold chunk is returned.

\paragraph{Soft@5.}

Soft@5 evaluates retrieval of semantically relevant supporting evidence that may not exactly match the manually annotated gold chunk.

\paragraph{Abstention Accuracy.}

For negative queries, abstention accuracy is defined as

\begin{equation}
\mathrm{Abstention}
=
\frac{N_{\mathrm{correct\ abstain}}}
     {N_{\mathrm{negative}}},
\end{equation}

where $N_{\mathrm{negative}}$ is the number of unsupported benchmark queries.

\section{Reproducibility Commands}
\label{app:repro}

The retrieval benchmarks are reproduced with the following commands. The retrieval and grounded-literature-access entry point is \texttt{chat.py}; the benchmark driver \texttt{evaluate\_jinst.py} runs all six retrieval configurations against the chunk-level gold benchmark and reports strict, paper-level, soft-gold, and abstention metrics.

\paragraph{Core 60-query benchmark.}

\begin{verbatim}
python evaluate_jinst.py \
  --benchmark data/eval/benchmark_jinst_gold_final.json \
  --chat ./chat.py \
  --configs bm25,dense,hybrid,graph,graph_path,agentic_graph \
  --top_k 10 \
  --candidate_k 80 \
  --generate \
  --out_dir data/eval/runs/jinst_eval_gold_final_rerank_v2
\end{verbatim}

\paragraph{Extension 56-query benchmark.}

\begin{verbatim}
python evaluate_jinst.py \
  --benchmark data/eval/benchmark_pixel_extension56_final_v1.json \
  --chat ./chat.py \
  --configs bm25,dense,hybrid,graph,graph_path,agentic_graph \
  --top_k 10 \
  --candidate_k 80 \
  --generate \
  --out_dir data/eval/runs/pixel_extension56_final_full_v2
\end{verbatim}

\paragraph{Single-query interactive retrieval and grounded response.}

\begin{verbatim}
python chat.py \
  --query "Time resolution of irradiated LGAD sensors" \
  --mode hybrid \
  --top_k 10 \
  --json
\end{verbatim}

\paragraph{Figure and table generation.}

\begin{verbatim}
python make_paper_graphs.py
\end{verbatim}

Each output directory contains the summary file, per-query metric file, and retrieved-evidence payloads used to audit the reported numbers:

\begin{verbatim}
summary.csv
per_query_metrics.csv
payloads/
\end{verbatim}

The reported core benchmark results correspond to

\begin{verbatim}
data/eval/runs/jinst_eval_gold_final_rerank_v2/
\end{verbatim}

and the reported extension benchmark results correspond to

\begin{verbatim}
data/eval/runs/pixel_extension56_final_full_v2/
\end{verbatim}

\bibliographystyle{JHEP}
\bibliography{references}

\end{document}